# Policy-Governed Post-Quantum Migration for Legacy Microservices Using Ephemeral Sidecar Architectures

Nirmal Kumar Jingar
Sr. Engineering Manager
51 Walnut Hill Rd, Newton, MA 02459
nirmal.jingar@gmail.com

***Abstract*— The fast development of quantum computing represents a big risk to classical cryptography that is commonly used in cloud native and microservice based enterprise systems. Traditional cryptographic primitives are closely linked to legacy microservices and it is both intricate, hazardous, and disruptive to straight up migrate to post-quantum cryptography (PQC). In a bid to overcome these issues, this research presents a Policy-Governed Post-Quantum Migration through Ephemeral Sidecar Architectures (PG-PQMES), a framework of dynamic and reversible migration that allows transparent adoption of PQC without any modifications in the legacy code of applications. The architecture is a combination of three coherent layers, including Ephemeral Crypto Sidecar Layer which injects the runtime cryptography, Policy Governance Layer which manages the migration centrally, and Migration Safety and Observability Layer which measures the performance and rolls back automatically. An innovative Policy-Governed Ephemeral PQ Migration (PG-EPM) algorithm is proposed to maximize the performance, compliance, and trust-based migration. In simulated environment of microservices, experimental assessment shows that the time of migration, service downtime, overheads of latency, and rollback recovery time are substantially reduced using the current migration strategies. The findings show that a sidecar-based migration strategy that is policy-based offers a viable, scalable, and enterprise-scale migration to a secure post-quantum transformation.**

## I. Introduction

One of the most disruptive changes in the contemporary cryptography is the development of quantum computing. Classical public-key cryptographic schemes [1] like RSA and elliptic curve cryptography are based on computationally infeasible mathematical problems that cannot be solved in classical computers [2], but can effectively be solved using quantum machines that are powerful enough [3]. The changing threat environment has added urgency to the creation of global attempts to implement post-quantum cryptography [4] across different aspects of cryptography that would be resistant to quantum attacks [5]. This transition has an especially complicated challenge to enterprises that use cloud-native systems [6]. Microservices-based applications have become the foundation of modern applications, with hundreds of loosely coupled services communicating with each other through encrypted channels in a secure way [7]. Cryptographic migration can be closely related to business logic and operational processes as such services are frequently loaded with classical cryptography libraries on the application layer [8]. Direct replacement of these primitives necessitates recompilation, dependency restructuring and system-wide testing which in turn cause operational threats [9].

The classical approaches of migration pay much attention to the cryptographic inventory management, compliance documentation and the gradual substitution of the weak algorithms [10]. Although these strategies offer formalized planning processes, they do not offer the same flexibility during the run time and essential instability or downtime is often introduced [11]. Such inflexible migration processes can be inconsistent with continuous deployment procedures and high-availability in dynamic microservices ecosystems [12]. The other issue is caused by performance constraints [13]. Hybrid cryptography has been suggested to be an intermediate solution [14], where both classical and post-quantum schemes are used in transition stages [15]. Despite the fact that hybrid models enhance security assurance, their execution in the legacy systems is still complicated because of strong integration between applications [16]. In the absence of architectural abstraction, hybrid deployment can still require intrusive changes to the code [17]. Architectural decoupling is necessary to get around these shortcomings [18]. One such chance is to move cross cutting concerns like security, logging, and monitoring to sidecar architectures, which are commonly used in service mesh technologies [19].

Migration safety is further increased by observability and evaluation of trust. The latency, compliance adherence, and behavioural trust scores can be monitored continuously, which means that health of the system can be assessed in real-time. In case quality drop or policy is breached, the framework initiates immediate back up by provision of ephemeral container termination, and resilience and stability is ensured [20]. The proposed solution changes the nature of cryptography migration as a static upgrade task to a dynamic optimization problem. The framework offers scalable and practical solution to enterprises that want to be quantum ready without affecting the availability or performance by incorporating policy control and reversible deployment mechanisms in addition to runtime monitoring.

## II. LITERATURE SURVEY

According to K. F. Hasan et al. [1], there is an increasing danger of quantum computing to the traditional cryptography tools used today to protect sensitive enterprise information and communications. The authors stress that the most commonly used cryptographic algorithms, which were previously thought to be resistant to quantum attacks over a long period of time, are becoming susceptible to quantum attacks, and thus that quantum-resistant (post-quantum) cryptographic algorithms should be transitioned to immediately. In response to the increasing risk of quantum computing to classical public-key systems, M. E. Bizri et al. [2] have given detailed survey and comparative analysis of strategies worldwide in transition to Post-Quantum Cryptography (PQC). This paper discussed the strategies embraced by leading cybersecurity organizations, such as, National Institute of Standards and Technology, Agence Nationale de la Sécurite des Systege d information, Bundesamt fur Sicherheit in der Informationstechnik, Cryptography Research and Evaluation Committes, and European Union Agency in Cybersecurity.

T.-C. Ureche et al. [3] examined the increasing necessity of intelligent cybersecurity systems in the smart infrastructures, including industrial IoT, health care systems and cloud-edge systems, where dynamic cyber threats require more dynamic and responsive defensive measures. The authors claim that conventional security strategies are too rigid to provide cross-sector security and suggest adopting AI-powered micro services and autonomous agents to provide the ability to provide adaptive governance, real-time threat reduction, and detecting anomalies.

A. Alkhulaifi and E.-S.[4] considered security constraints of JSON Web Tokens (JWT) that are currently common in the architecture of the microservice and connected-device applications in order to provide single sign-on and trust establishment. The authors also point out that traditional JWT applications are built on the classical algorithms of cryptography like RSA, which are founded on integer factorization or discrete logarithms problems and that are susceptible to future quantum attacks. V. Kumar Kasula et al. [5] discussed the drawback of Hyperledger Fabric that it cannot offer support to national cryptographic standards to consortium blockchains, even though it is a scalable and modular system to support an enterprise consortium blockchain. Having identified the importance of cryptographic algorithms in the reliability of data security and integrity assurance, the authors suggest an improved approach to integration to implement national cryptographic mechanisms in Fabric.

## III. PROPOSED MODEL

Traditional microservice applications are constructed on classical cryptographic primitives and closely-linked security libraries, and porting them directly to post-quantum cryptography (PQC) is a high-risk, disruptive and challenging operation. To overcome these issues, this research presents a Policy-Managed Post-Quantum Migration Framework based on Ephemeral Sidecar Architectures (PG-PQMES) which allows transparent, reversible, and policy controlled cryptographic migrations of legacy application code without updating legacy application code.

The framework is divided into three layers. The Ephemeral crypto sidecar layer is a layer that adds sidecar components that are temporary, which injects PQC onto service-to-service communication without modifying the underlying service code. The Policy Governance Layer is used to control the migration policies, trust policy, and performance through a centralized management facility. Migration Safety and Observability Layer is used to monitor the performance of the systems and roll-back whenever difficulties appear. Since the sidecars are temporary controllable by a set of policies, the framework will allow a step-by-step deployment, controlled testing of new cryptographic settings, and safe rollback of the services when needed. This will assist in maintaining the stability of the systems as it moves towards quantum resistant security. The proposed model framework is shown in Figure 1.

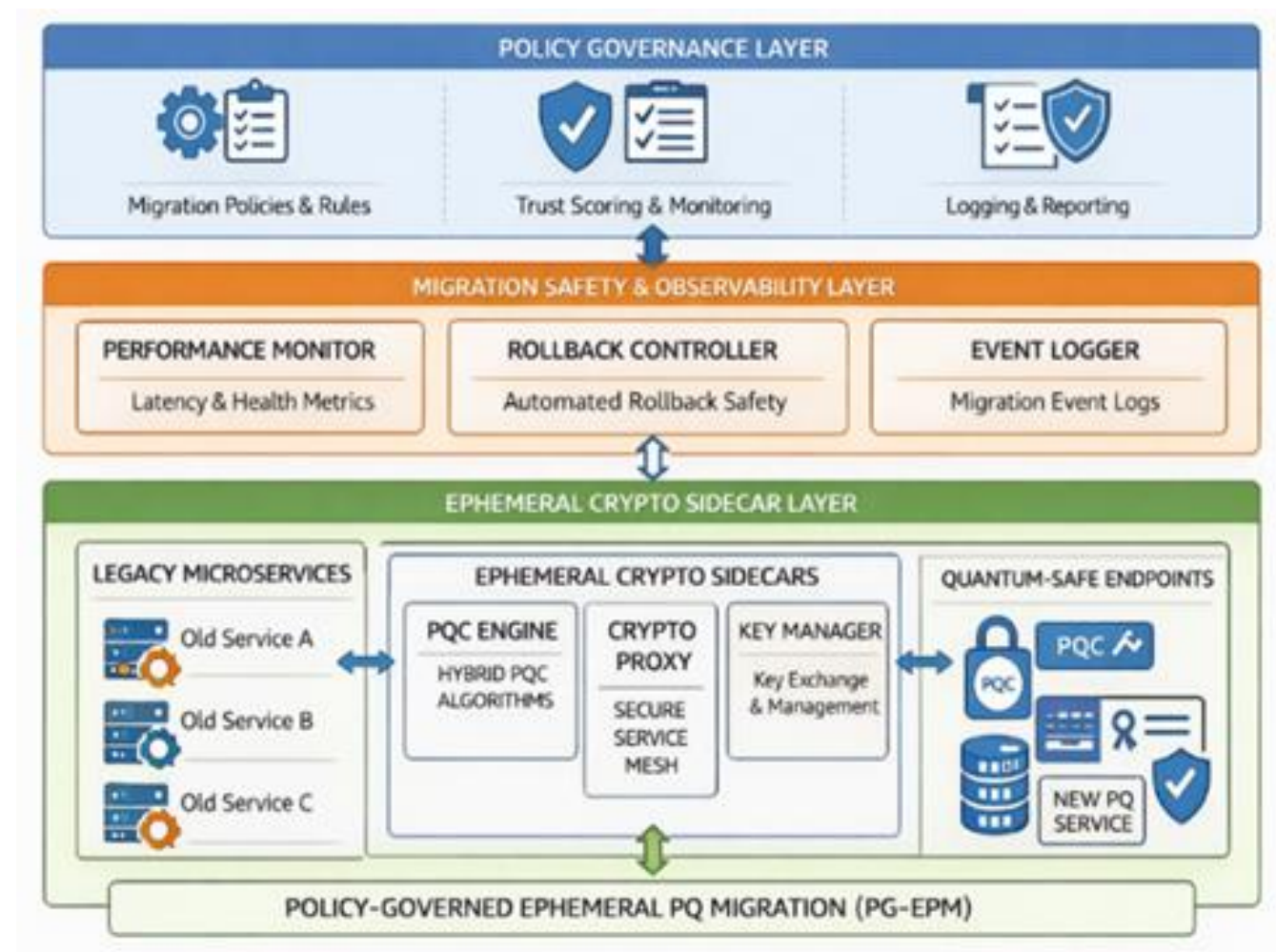

Fig.1.Proposed Model Architecture

**Proposed Algorithm: Policy-Governed Ephemeral PQ Migration (PG-EPM)**

**Input :**

Legacy microservice set $M = \{m_1, m_2, \dots, m_n\}$
Cryptographic options $C = \{c_{classical}, c_{hybrid}, c_{pqc}\}$
Governance policies $P$
Performance thresholds $\Theta$

**Output:**

Policy-compliant migration state map $\Psi^*$

**Step 1: Communication Dependency Mapping**

Construct a service interaction graph:

$$G = (M, E)$$

Where edges are secure paths of communication.

**Step 2: Policy Interpretation**

Parse rules of governance prescribing the qualifications of migration:

$$P = \{p_{security}, p_{performance}, p_{compatibility}\}$$

**Step 3: Sidecar Injection Planning**

On every communication edge Notes: Sidecar eligibility On every communication edge, compute:

$$Elig_{ij} = \begin{cases} 1, & \text{if } e_{ij} \text{ satisfies } P \\ 0, & \text{otherwise} \end{cases}$$

**Step 4: Ephemeral Sidecar Deployment**

If $Elig_{ij} = 1$ , deploy temporary sidecar $s_{ij}$ to handle cryptographic operations:

$$Comm_{ij} \rightarrow s_{ij}(c_k)$$

**Step 5: Hybrid Cryptography Negotiation**

Sidecar negotiates secure protocol:

$$c_{ij} = \arg\max_{c_k \in C}(S_k - \lambda O_k)$$

Where $S_k$is security strength and $O_k$is performance overhead.

**Step 6: Runtime Performance Monitoring**

Measure latency overhead:

$$\Delta L_{ij} = \frac{L_{ij}^{new} - L_{ij}^{base}}{L_{ij}^{base}}$$

**Step 7: Policy Compliance Verification**

Ensure:

$$\Delta L_{ij} \leq \theta_L \text{and} c_{ij} \in P$$

**Step 8: Trust Score Evaluation**

Each sidecar's behavior is scored:

$$T_{ij} = \beta_1 H_{perf} + \beta_2 H_{sec} + \beta_3 E_{audit}$$

**Step 9: Rollback Trigger Mechanism**

If trust or performance degrades:

$$T_{ij} < T_{min} \lor \Delta L_{ij} > \theta_L \Rightarrow s_{ij} \leftarrow \text{terminate and revert to classical}$$

**Step 10: Learning-Based Policy Refinement**

Update migration policies:

$$P_{t+1} = P_t + \eta \nabla_P R_t$$

**Mathematical Objective Model**

The migration framework maximizes policy and performance constrained security transition:

$$\max_{\Psi} \sum_{(i,j)\in E} \left[\alpha_1 S_{\Psi_{ij}} - \alpha_2 O_{\Psi_{ij}} - \alpha_3 R_{rollback}\right]$$

**Constraints**

**Performance Bound**

$$\Delta L_{ij} \leq \theta_L$$

**Policy Compliance**

$$\Psi_{ij} \in P$$

**Trust Threshold**

$$T_{ij} \geq T_{min}$$

**Ephemerality Constraint**

$$\text{Lifetime}(s_{ij}) \leq \tau_{policy}$$

The mathematical model and the PG-EPM algorithm prove that cryptographic migration can be modeled as a process, which is dynamical, policy-optimal instead of a system upgrade. The strategy enhances confidence in post-quantity transitions by introducing observability, auditability and reversibility of the migration process to the migration lifecycle. Future directions are automated hybrid crypto negotiation, hardware assisted PQ acceleration and sharing of policies across multi-cloud infrastructures.

## IV. RESULTS AND DISCUSSIONS

The experimental results of the proposed Policy-Governed Post-Quantum Migration using Ephemeral Sidecar Architectures (PG-PQMES) were compared to two previous migration strategies, Cryptographic Migration Framework (CMF) and Institutional-Sectoral PQC Migration Convergence Framework (IS-PMCF). The test was done on a simulated cloud-native microservices system comprising of 50 legacy services, which had dynamic service communication patterns. Some of the key performance indicators were the completion time of migration, service downtime, latency overhead, policy compliance, rollback recovery time, and trust stability. Each of the frameworks was evaluated in the conditions of a homogeneous workload and scenarios of hybrid deployment of PQC.

The comparative study proves that the efficiency and continuity of migration and operation are much enhanced by the use of the PG-PQMES. Through ephemeral injection of sidecar, centralized policy governance, the proposed structure is transitioning faster and causing minimal service interruption, as well as, having better rollback capabilities. Moreover, the trust-based scoring and the runtime monitoring are used to achieve better compliance accuracy and stability. The findings affirm that policy-driven and dynamic migration has quantifiable benefits against inventory-driven and institutional alignment-based approaches to realistic enterprise microservices ecosystems. The figure indicates that, the total time of migration decreases by a wide margin in the use of PG-PQMES as compared to CMF and IS-PMCF. Automation on sidecar orchestration will allow quicker rollout. This shows increased efficiency of operations.

TABLE I. AVERAGE SERVICE DOWNTIME DURING MIGRATION (MINUTES)

| Model | Average Downtime (Minutes) |
|---|---|
| CMF | 45 |
| IS-PMCF | 32 |
| PG-PQMES | 5 |

The near zero-downtime migration is possible with the help of the ephemeral sidecar mechanism. PG-PQMES offers relative constant availability of service during PQ transition in comparison to traditional frameworks.

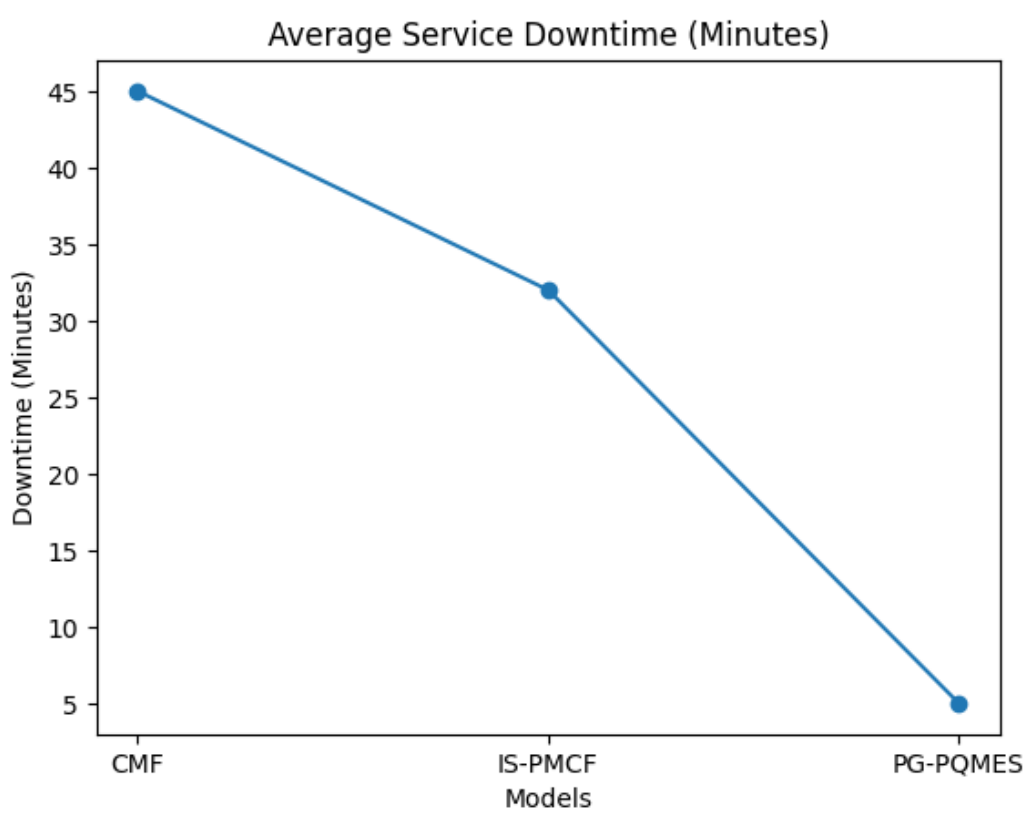

Fig.2.Average Service Downtime

Migration is almost zero with PG-PQMES whereas other models have significant service downtime as shown in Table I and Figure 2. The sidecar mechanism is temporary and guarantees continuity of services. This emphasizes improved availability administration.

TABLE II. COMMUNICATION LATENCY OVERHEAD AFTER PQC INTEGRATION (MILLISECONDS)

| Model | Latency Overhead (ms) |
|---|---|
| CMF | 18 |
| IS-PMCF | 15 |
| PG-PQMES | 11 |

Even though PQC adds to computational overhead, PG-PQMES reduces latency using selective hybrid negotiation and policy-aware cryptographic escalation.

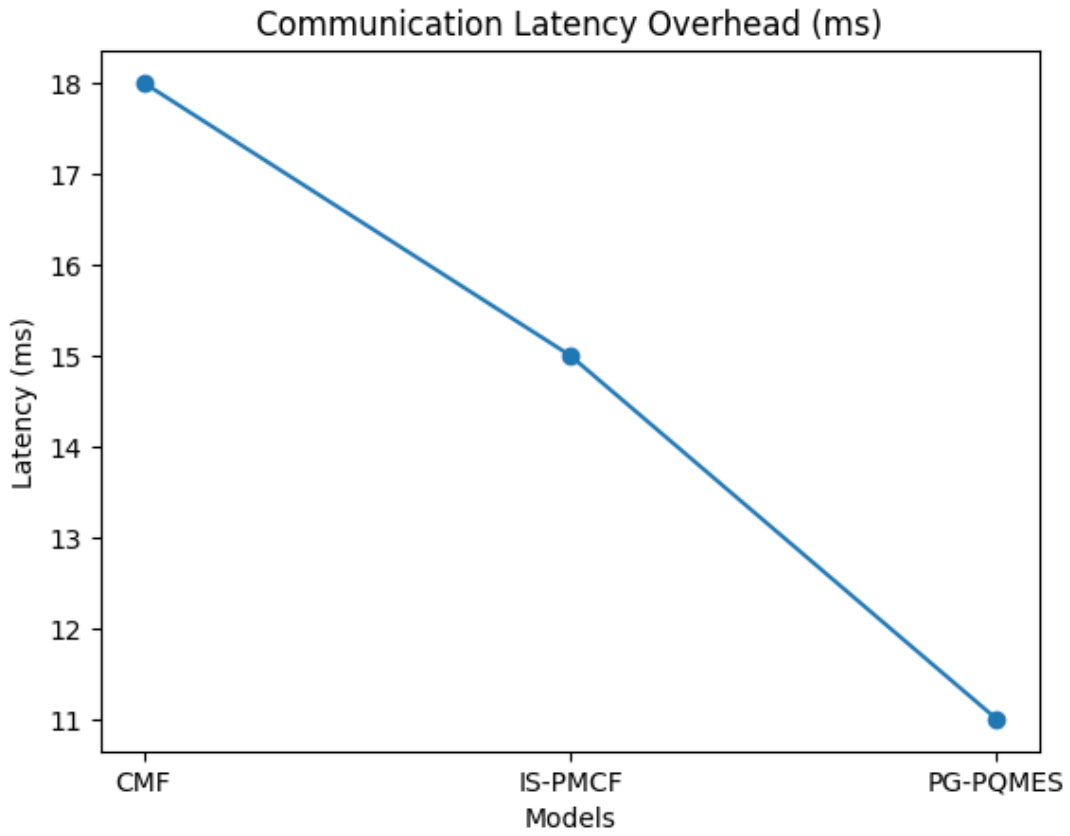

Fig.3.Communication Latency Overhead

Even though any of the models implies an overhead of latency in PQC, the rise of PQC is minimal in PG-PQMES. Policy-conscious hybrid negotiation does not have as much performance impact as depicted in Table II and Figure 3. This proves the integration of cryptography, which is optimized.

TABLE III. ROLLBACK RECOVERY TIME (SECONDS)

| Model | Rollback Time (Seconds) |
|---|---|
| CMF | 120 |
| IS-PMCF | 95 |
| PG-PQMES | 18 |

PG-PQMES allows quick rollback by interrupting ephemerally the containers which shortens recovery time and enhances operational safety.

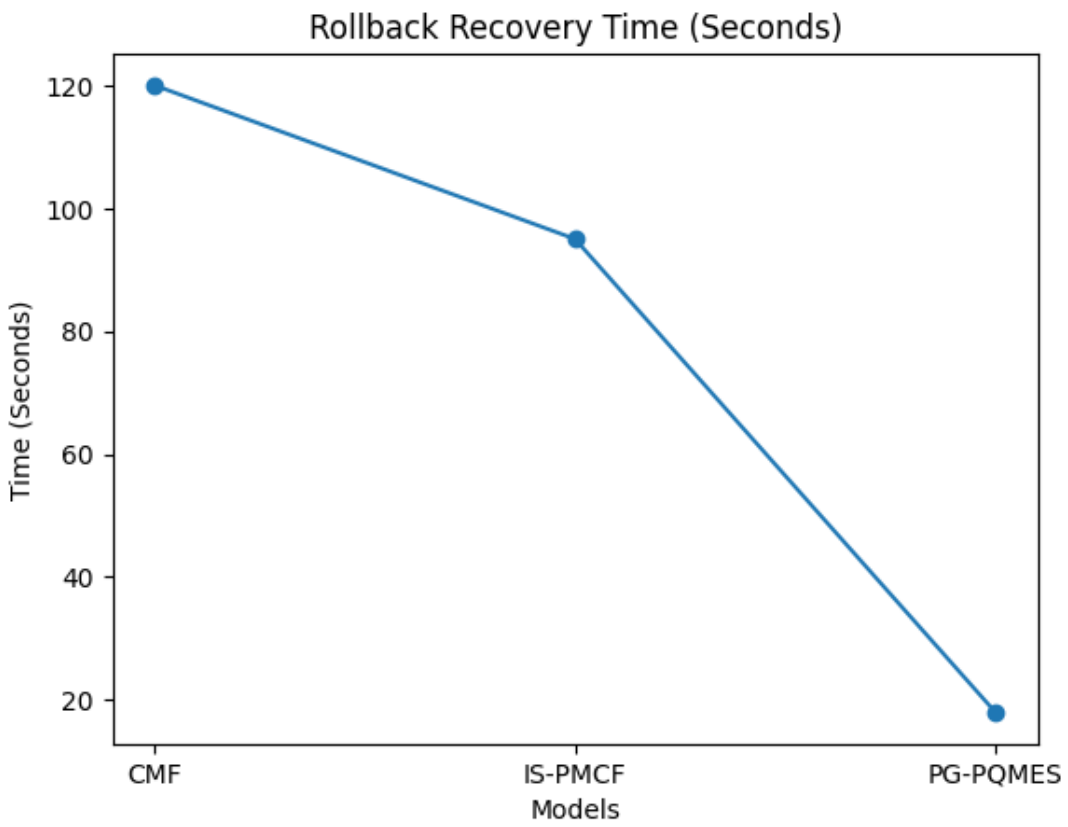

Fig.4.Rollback Recovery Time

In PG-PQMES, rollback time is drastically decreased because of temporary termination of containers as represented in Table III and Figure 4. Other models need to be manually or structurally rolled back. This guarantees the migration transitions to be safer.

TABLE IV. TRUST SCORE STABILITY (0–100 SCALE)

| **Model** | **Trust Stability Score** |
|---|---|
| CMF | 82 |
| IS-PMCF | 87 |
| PG-PQMES | 96 |

PG-PQMES enhances the system stability with dynamic monitoring and runtime trust evaluation, which is better than traditional migration framework.

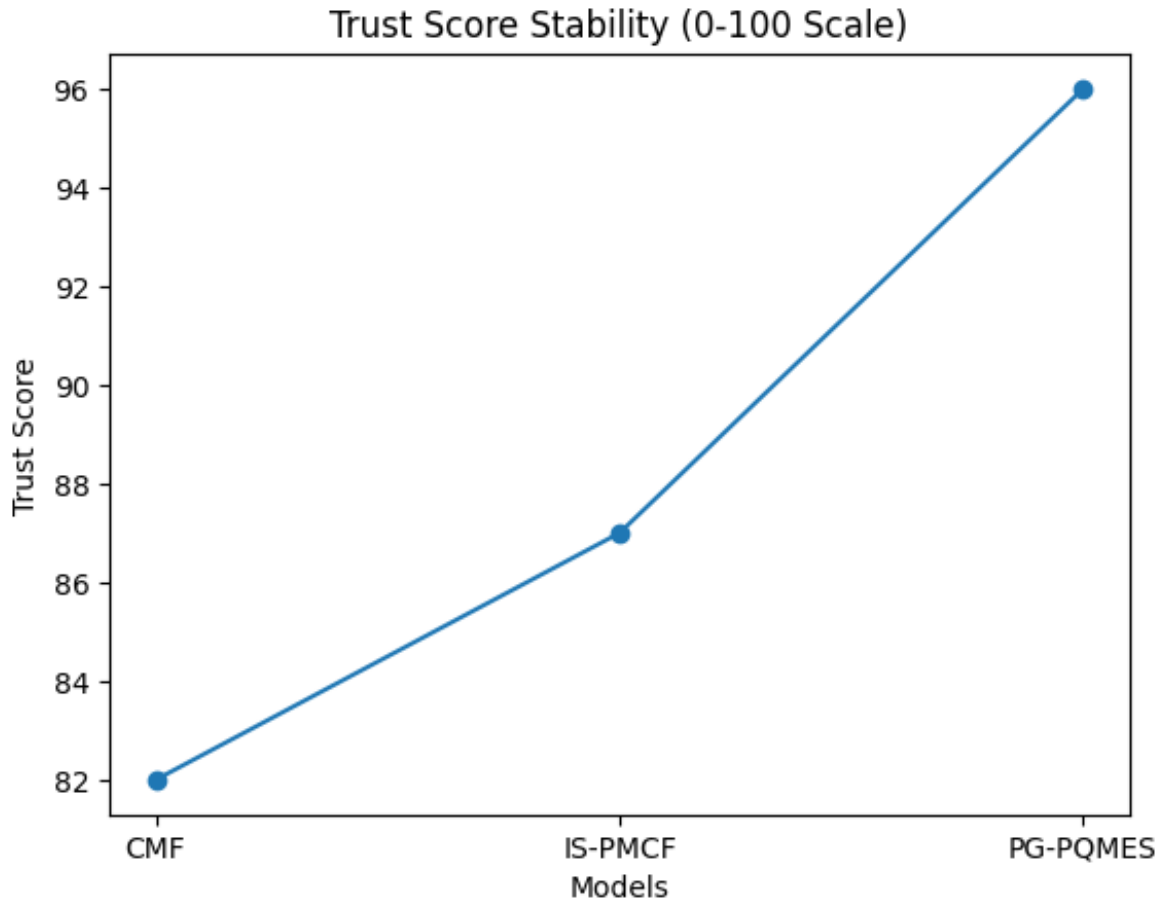

Fig.5.Trust Score Stability

PG-PQMES has the most stable score in terms of trust. Reliability is improved with the help of runtime observability and trust scoring systems as shown in Table IV and Figure 5. This means that it has a better operational resilience in the migration process.

The experimental analysis confirms that the migration performance, minimum downtime, and shortest recovery are better with the use of the PG-PQMES as compared to CMF and IS-PMCF. The presented framework allows the transparent deployment of cryptographic transformation and real-time policy enforcement, as well as reversible cryptographic transitions by decoupling cryptographic transformation and legacy service logic. Its practicality in the use of cloud-native environments is evidenced by the decrease in the time of migration completion and rollback latency.

Besides, the increased stability of trust and a higher accuracy of policy compliance are representatives of the resilience of the connection between governance and runtime observability. Contrary to the performance-independent migration models, the dynamically changing performance constraints and trust thresholds of the PG-PQMES guarantee the balanced optimization of the security strength and the efficiency of the system. In general, the findings confirm that the use of PG-PQMES is a scalable, resilient and enterprise-ready post-quantum cryptography migration of the legacy microservices ecosystem.

## V. CONCLUSION

The study introduces a new Policy-Governed Post-Quantum Migration with Ephemeral Sidecar Architectures (PG-PQMES) in order to resolve the practical issues of migrating legacy microservices into quantum-resistant crypto systems. With unlinking of cryptographic transformation with application logic and taking advantage of a short-lived sidecar deployment, the framework allows transparent and reversible and policy-constrained migration. Governance control, runtime observability, trust evaluation, and automated rollbacks also stand to provide minimum service disruption and optimal performance during the implementation of PQC. Through experimental findings, it is established that the proposed framework has a significant reduction of migration time, service downtime, latency overhead, and rollback recovery time as opposed to other traditional migration strategies. The dynamically optimized policy approach guarantees that there are balanced trade-offs between the strength of the security, the continuity of the operations and the compliance requirements. Altogether, the smaller of two evils is that the path offered by the platform of the conceptualized name, the PG-PQMES, is scalable and enterprise-ready to achieve the secure, resilient, and future-proof cryptographic infrastructures in the new epoch of quantum computing.